\documentclass[oneside,11pt]{article}

\usepackage[small,it]{caption}
\usepackage{url,mathrsfs,algorithm}
\usepackage{amsmath,wrapfig,color}
\usepackage{times}
\usepackage{amsfonts}
\usepackage{graphicx}
\usepackage{subfigure}
\newtheorem{theorem}{Theorem}
\newtheorem{lemma}{Lemma}

\begin{document}

\title{\huge Compressed Sensing  with Very Sparse Gaussian Random Projections}

\author{ \textbf{Ping Li} \\
         Department of Statistics and Biostatistics\\
         Department of Computer Science \\
       Rutgers University\\
         Piscataway, NJ 08854, USA\\
      \texttt{pingli@stat.rutgers.edu}
       \and
 \textbf{Cun-Hui Zhang} \\
         Department of Statistics and Biostatistics\\
       Rutgers University\\
         Piscataway, NJ 08854, USA\\
         \texttt{cunhui@stat.rutgers.edu}
        }
\date{}
\maketitle

\begin{abstract}

\noindent We study the use of {\em very sparse random projections} for compressed sensing (sparse signal recovery) when the  signal  entries can be either positive or negative. In our setting, the entries of a Gaussian design matrix are randomly sparsified so that only a very small fraction of the entries are nonzero. Our proposed decoding algorithm is simple and efficient in that the major cost is one linear scan of the coordinates. We have developed two estimators: (i) the {\em tie estimator}, and (ii) the {\em absolute minimum estimator}. Using only the tie estimator, we are able to recover a $K$-sparse signal of length $N$ using  $1.551 eK \log K/\delta$ measurements (where $\delta\leq 0.05$ is the confidence). Using only the absolute minimum estimator, we can detect the support of the signal using $eK\log N/\delta$ measurements. For a particular coordinate, the absolute minimum estimator requires fewer measurements (i.e., with a constant $e$ instead of $1.551e$). Thus, the two estimators can be combined to form an even more practical decoding framework.\\

\noindent Prior studies have shown that existing one-scan (or roughly one-scan) recovery algorithms using sparse matrices would  require substantially more (e.g., one order of magnitude) measurements than L1 decoding by linear programming, when the nonzero entries of signals can be either negative or positive.  In this paper, following a known experimental setup~\cite{Url:wiki_sparse}\footnote{\scriptsize\url{http://groups.csail.mit.edu/toc/sparse/wiki/index.php?title=Sparse_Recovery_Experiments}}, we show that, at the same number of measurements, the recovery accuracies   of our proposed method are (at least) similar to   the standard L1 decoding.

\end{abstract}

\newpage

%\vspace{-0.1in}
\section{Introduction }

{\em Compressed Sensing (CS)}~\cite{Article:Donoho_CS_JIT06,Article:Candes_Robust_JIT06} has become an important and popular topic in several fields, including Computer Science, Engineering, Applied Mathematics, and Statistics.  The goal of {compressed sensing} is to recover a sparse signal $\mathbf{x}\in\mathbb{R}^{1\times N}$ from a small number of non-adaptive linear measurements $\mathbf{y} = \mathbf{xS}$, where $\mathbf{S}\in\mathbb{R}^{N\times M}$ is the ``design'' matrix (or ``sensing'' matrix). Typically, the signal $\mathbf{x}$ is assumed to be $K$-sparse (i.e., $K$ nonzero entries) and neither the magnitudes nor locations of the nonzero coordinates are known. Many streaming/database applications can be naturally formulated as compressed sensing problems~\cite{Article:Charikar_2004,Article:Cormode_05,Article:Muthukrishnan_05} (even before the name ``compressed sensing'' was proposed).  The idea of compressed sensing may be traced back to many prior papers, for example~\cite{Article:Stark_89,Article:Huo_JIT01}.

In the  literature of compressed sensing, entries of the design matrix $\mathbf{S}$ are often sampled i.i.d. from a Gaussian distribution (or Gaussian-like distribution, e.g., a distribution with a finite second moment). Well-known recovery algorithms are often based on linear programming (LP) (e.g., {\em basis pursuit}~\cite{Article:Chen98} or L1 decoding) or greedy  methods such  as {orthogonal matching pursuit (OMP)}~\cite{Proc:Pati93,Article:Mallat93,Article:Zhang_RIP11,Article:Tropp_JIT04}. In general, L1 decoding is computationally  expensive. OMP is often   more efficient than L1 decoding but   it  can still be expensive especially when $K$ is large.

\subsection{Compressed Sensing with Very Sparse Random Projections }

The process of collecting measurements, i.e., $\mathbf{y} = \mathbf{xS}$, is often called ``random projections''. \cite{Proc:Li_Hastie_Church_KDD06} studied the idea of ``very sparse random projections'' by randomly sparsifying  the sensing matrix $\mathbf{S}$ so that only a very small fraction of the entries can be nonzero. In this paper, we will continue to investigate on the idea of very sparse random projections in the context of compressed sensing.

Our work is   related to  ``sparse recovery with sparse matrices''~\cite{Proc:SMP08,Proc:Berinde_CCC08,Article:Gilbert_IEEE10,Article:Omidiran_JMLR10}, for example, the SMP ({\em Sparse Matching Pursuit}) algorithm~\cite{Proc:SMP08}. There is a  nice  well-known wiki page~\cite{Url:wiki_sparse}, which summarizes the comparisons of L1 decoding with count-min sketch~\cite{Article:Cormode_05} and SMP. Their results have shown that, in order to achieve similar recovery accuracies,  count-min sketch needs about $10$ to 15 times more measurements than L1 decoding and SMP  needs about half of the measurements of count-min sketch.

In comparison, our experimental section (e.g., Figure~\ref{fig_g1s1}) demonstrates that the proposed method can  be as accurate as (or even more accurate than) L1 decoding, at the same number of measurements. The major cost of our method is one linear scan of the coordinates, like count-min sketch.

\subsection{Linear Measurements from Sparse Projections}

In this paper, our  procedure for compressed sensing first collects $M$ non-adaptive linear measurements
\begin{align}
y_j = \sum_{i=1}^N x_i \left[s_{ij}r_{ij}\right],\hspace{0.5in} j = 1, 2, ..., M
\end{align}
Here, $s_{ij}$ is the $(i,j)$-th entry of the design matrix with $s_{ij}\sim N(0,1)$ i.i.d.  Instead of using a dense design matrix, we randomly sparsify $(1-\gamma)$-fraction of the entries of the design matrix to be zero, i.e.,
\begin{align}
r_{ij} = \left\{\begin{array}{ll}
1 & \text{ with prob. } \gamma \\
0 & \text{ with prob. } 1-\gamma
\end{array}\right.\ \ \ i.i.d.
\end{align}
And any $s_{ij}$ and  $r_{ij}$ are also independent.\\

Our proposed decoding scheme utilizes two simple estimators: (i) the {\em tie estimator} and (ii) the {\em absolute minimum estimator}. For convenience, we will theoretically analyze them separately. In practice,  these two estimators should be combined to form a powerful decoding framework.

\subsection{The Tie Estimator}

The tie estimator is developed based on the following interesting observation on  the {\em ratio statistics} $\frac{y_j}{s_{ij}r_{ij}}$.  Conditional on $r_{ij}=1$, we can write
\begin{align}
\left.\frac{y_j}{s_{ij}r_{ij}}\right|_{r_{ij}=1}  = \frac{\sum_{t=1}^Nx_ts_{tj}r_{tj}}{s_{ij}}=x_i + \frac{\sum_{t\neq i}^Nx_ts_{tj}r_{tj}}{s_{ij}} = x_i + \left(\eta_{ij}\right)^{1/2} \frac{S_2}{S_1}
\end{align}
where $S_1, S_2\sim N(0,1)$, i.i.d., and
\begin{align}\label{eqn_eta}
\eta_{ij} = \sum_{t\neq i}^N \left|x_t r_{tj}\right|^2= \sum_{t\neq i}^N \left|x_t\right|^2 r_{tj}
\end{align}
Note that $\eta_{ij}$ has certain probability of being zero. If $\eta_{ij} = 0$, then $\left.\frac{y_j}{s_{ij}r_{ij}}\right|_{r_{ij}=1} = x_i$. Thus, given $M$ measurements, if  $\eta_{ij} =0$ happens (at least) \textbf{twice} (i.e., a \textbf{tie} occurs), we can exactly identify the value $x_i$. This is the key observation which  motivates  our proposal of the tie estimator. \\

Another  key observation is that, if $x_i=0$, then we will not see a nonzero tie (i.e., the probability of nonzero tie is 0). This is due to the fact that we use a Gaussian design matrix, which excludes unwanted ties. It is also clear that the Gaussian assumption is not needed, as long as $s_{ij}$ follows from a continuous distribution. In this paper we focus on Gaussian design because it makes some detailed analysis easier.

\subsection{The Absolute Minimum Estimator}

It turns out that, if we just need to detect whether $x_i=0$, the task is easier than estimating the value of $x_i$, for a particular coordinate $i$. Given $M$ measurements, if $\eta_{ij} = 0$ happens (at least) \textbf{once}, we will be able to determine whether $x_i=0$.  Note that unlike the tie estimator, this estimator will generate ``false positives''. In other words, if we cannot be certain that $x_i=0$, then it is still possible that $x_i=0$ indeed.

From the practical perspective, at a particular coordinate $i$, it is preferable to first detect whether $x_i=0$ because that would require fewer measurements than using the tie estimator.  Later in the paper, we can see that the performance can be potentially further improved by a more general estimator, i.e., the  so-called  {\em absolute minimum estimator}:
\begin{align}
\hat{x}_{i,min,\gamma} = z_{i,t}, \ \ \text{ where } t = \underset{1\leq j\leq M}{\text{argmin}}\  |z_{i,j}|,\hspace{0.2in} z_{ij} = \frac{y_j}{s_{ij}r_{ij}}
\end{align}
We will also introduce a threshold $\epsilon$ and provide a theoretical analysis of the event  $\hat{x}_{i,min,\gamma}\geq \epsilon$. When $\epsilon=0$, it becomes the ``zero-detection'' algorithm. Our analysis will show that by using $\epsilon>0$ we can better exploit the prior knowledge we have about the signal and hence improve the accuracy.

\subsection{The Practical Procedure}

We will separately analyze the tie estimator and the absolute minimum estimator, for the  convenience of theoretical analysis. However, we recommend a mixed procedure. That is, we first run the absolute minimum estimator in one scan of the coordinates, $i=1$ to $N$. Then we run the tie estimator only on those coordinates which are possibly not zero. Recall that the absolute minimum estimator may generate false positives.

As an option, we can iterate this process for several rounds. After one iteration (i.e., the absolute minimum estimator followed by the tie estimator), there might be a set of coordinates for which we cannot decide their values. We can compute the residuals and use them as the measurements for the next iteration. Typically, a few (e.g., 3 or 4) iterations are sufficient and the major computational cost is computing the absolute minimum estimator in the very first iteration.

\section{Analysis of the Absolute Minimum Estimator}

The important task is to analyze the false positive probability: $\mathbf{Pr}\left(|\hat{x}_{i,min,\gamma}|>\epsilon, x_i=0\right)$ for some chosen threshold $\epsilon >0$. Later we will see that $\epsilon$ is irrelevant if we only care about the worst case.

Recall that, conditional on $r_{ij}=1$, we can express $\frac{y_j}{s_{ij}r_{ij}} = x_i + \left(\eta_{ij}\right)^{1/2} \frac{S_2}{S_1}$, where $S_1, S_2 \sim N(0,1)$ i.i.d. and $\eta_{ij}$ is defined in (\ref{eqn_eta}). It is known that $S_2/S_1$ follows the standard Cauchy distribution. Therefore,
\begin{align}
\mathbf{Pr}\left(\left|\frac{S_2}{S_1}\right|\leq t\right) = \frac{2}{\pi}\tan^{-1}(t), \ \ t>0
\end{align}

We are ready to present  the Lemma about the false positive probability, including a practically useful data-dependent bound, as well as a data-independent bound (which is convenient for worst-case analysis).

\subsection{The False Positive Probability}

\begin{lemma}\label{lem_Pr_fp}
\textbf{Data-dependent bound:}
\begin{align}\label{eqn_fp_exp}
\mathbf{Pr}\left(\left|\hat{x}_{i,min,\gamma}\right|>  \epsilon, x_i = 0\right)
=&\left[1- \gamma E\left\{\frac{2}{\pi}\tan^{-1}\left(\frac{\epsilon}{\eta_{ij}^{1/2}}\right)\right\}\right]^M\\\label{eqn_fp_bound}
\leq&\left[1- \gamma \left\{\frac{2}{\pi}\tan^{-1}\left\{\frac{{\epsilon}}{\sqrt{\gamma\sum_{t} x_{t}^2}}\right\}\right\}\right]^M
\end{align}
\textbf{Data-independent (worst case) bound:}
\begin{align}\label{eqn_fp_worst}
\mathbf{Pr}\left(\left|\hat{x}_{i,min,\gamma}\right|>  \epsilon, x_i = 0\right)
\leq\left[1-\gamma\left(1-\gamma\right)^K\right]^M
\end{align}

\end{lemma}

\noindent\textbf{Remark:}\hspace{0.2in} The data-dependent bound (\ref{eqn_fp_exp}) and (\ref{eqn_fp_bound}) can be numerically evaluated if we have information about the data. The bound will help us understand why empirically the performance of our proposed algorithm is substantially better than the worst-case bound. On the other hand, the worst case bound (\ref{eqn_fp_worst}) is convenient for theoretical analysis. In fact, it directly leads to the $eK\log N$ complexity bound.\\

\noindent\textbf{Proof of Lemma~\ref{lem_Pr_fp}}:  \hspace{0.2in} For convenience, we define  the set $T_i = \{j,\ 1\leq j\leq M, \ r_{ij}= 1\}$.
\begin{align}\notag
&\mathbf{Pr}\left(\left|\hat{x}_{i,min,\gamma}\right|>  \epsilon, x_i = 0\right)
=E\left(\mathbf{Pr}\left(\left|\frac{y_j}{s_{ij}}\right| >  \epsilon,\ x_i = 0, \  j\in T_i|T_i\right)\right)\\\notag
=&E\prod_{j\in T_i} \left[ \mathbf{Pr}\left(\left|\frac{S_2}{S_1}\right| >\frac{\epsilon}{\eta_{ij}^{1/2}}, x_i = 0\right)\right]
=E\prod_{j\in T_i}\left[1-\frac{2}{\pi}\tan^{-1}\left(\frac{\epsilon}{\eta_{ij}^{1/2}}\right)\right]\\\notag
=&E\left\{\left[1-E\left\{\frac{2}{\pi}\tan^{-1}\left(\frac{\epsilon}{\eta_{ij}^{1/2}}\right) \right\}\right]^{|T_i|}\right\}\\\notag
=&\left[1-\gamma + \gamma\left\{1-E\left\{\frac{2}{\pi}\tan^{-1}\left(\frac{\epsilon}{\eta_{ij}^{1/2}}\right)\right\}\right\}\right]^M\\\notag
=&\left[1- \gamma E\left\{\frac{2}{\pi}\tan^{-1}\left(\frac{\epsilon}{\eta_{ij}^{1/2}}\right)\right\}\right]^M
\end{align}

By noticing that $f(x) = \tan^{-1}\frac{a}{\sqrt{x}}$, (where $a>0$), is a convex function of $x>0$, we can obtain an upper bound by using Jensen's inequality.
\begin{align}\notag
&\mathbf{Pr}\left(\left|\hat{x}_{i,min,\gamma}\right|> \epsilon, \ x_i=0\right)\\\notag
=&\left[1- \gamma E\left\{\frac{2}{\pi}\tan^{-1}\left(\frac{\epsilon}{\eta_{ij}^{1/2}}\right)\right\}\right]^M\\\notag
\leq&\left[1- \gamma \left\{\frac{2}{\pi}\tan^{-1}\left(\frac{{\epsilon}}{\left(E\eta_{ij}\right)^{1/2}}\right)\right\}\right]^M \hspace{0.3in} (\text{Jensen's Inequality})\\\notag
=&\left[1- \gamma \left\{\frac{2}{\pi}\tan^{-1}\left\{\frac{{\epsilon}}{\left(\gamma\sum_{t\neq i} x_{t}^2\right)^{1/2}}\right\}\right\}\right]^M\\\notag
=&\left[1- \gamma \left\{\frac{2}{\pi}\tan^{-1}\left\{\frac{{\epsilon}}{\sqrt{\gamma\sum_{t} x_{t}^2}}\right\}\right\}\right]^M
\end{align}

We can further obtain a worst case bound as follows. Note that $\eta_{ij}$ has some mass at 0.
\begin{align}\notag
&\mathbf{Pr}\left(\left|\hat{x}_{i,min,\gamma}\right|> \epsilon, \ x_i=0\right)\\\notag
=&\left[1- \gamma E\left\{\frac{2}{\pi}\tan^{-1}\left(\frac{\epsilon}{\eta_{ij}^{1/2}}\right)\right\}\right]^M\\\notag
\leq&\left[1- \gamma \left\{\frac{2}{\pi}\tan^{-1}\left(\frac{\epsilon}{\textbf{0}}\right)\right\}\mathbf{Pr}\left(\eta_{ij}=0\right)\right]^M\\\notag
=&\left[1-\gamma\left(1-\gamma\right)^K\right]^M
\end{align}
$\hfill\Box$

\vspace{-0.1in}
\subsection{The False Negative Probability}
It is also necessary to control the false negative probability: $\mathbf{Pr}\left(\left|\hat{x}_{i,min,\gamma}\right|\leq  \epsilon, \ x_i\neq 0\right)$.

\begin{lemma}\label{lem_Pr_fn}
\begin{align}\notag
&\mathbf{Pr}\left(\left|\hat{x}_{i,min,\gamma}\right|\leq \epsilon, \ x_i\neq 0\right)\\\label{eqn_fn_exp}
=&1-\left[1-\gamma E\left\{\frac{1}{\pi}\tan^{-1}\left(\frac{\epsilon+x_i}{\eta_{ij}^{1/2}}\right)-\frac{1}{\pi}\tan^{-1}\left(\frac{x_i-\epsilon}{\eta_{ij}^{1/2}}\right)\right\}\right]^M\\\label{eqn_fn_bound}
\leq&1-\left[1-\frac{2}{\pi}\gamma\tan^{-1}\epsilon\right]^M
\end{align}\end{lemma}
\noindent\textbf{Remark:}\hspace{0.2in} Again, if we know information about the data, we might be able to numerically evaluate the exact false negative probability (\ref{eqn_fn_exp}). The (loose) upper bound (\ref{eqn_fn_bound}) is also insightful because it means this probability $\rightarrow0$ if $\epsilon\rightarrow0$. Note that in Lemma~\ref{lem_Pr_fp}, the worst case bound is actually independent of $\epsilon$. This implies that, if we only care about the worst case performance, we do not have to worry about the false positive probability since we can always choose $\epsilon\rightarrow0$.\\

\noindent \textbf{Proof of Lemma~\ref{lem_Pr_fn}:}
\begin{align}\notag
&\mathbf{Pr}\left(\left|\hat{x}_{i,min,\gamma}\right|\leq   \epsilon, x_i \neq 0\right)\\\notag
=&1-\mathbf{Pr}\left(\left|\hat{x}_{i,min,\gamma}\right|>   \epsilon, x_i \neq 0\right)\\\notag
=&1-E\left(\mathbf{Pr}\left(\left|\frac{y_j}{s_{ij}}\right| >  \epsilon,\ x_i \neq 0, \  j\in T_i|T_i\right)\right)\\\notag
=&1-E\prod_{j\in T_i} \left[ \mathbf{Pr}\left(\left|x_i + \eta_{ij}^{1/2}\frac{S_2}{S_1}\right| >\epsilon, x_i \neq 0\right)\right]\\\notag
=&1-E\prod_{j\in T_i}\left[1-\frac{1}{\pi}\tan^{-1}\left(\frac{\epsilon-x_i}{\eta_{ij}^{1/2}}\right) -\frac{1}{\pi}\tan^{-1}\left(\frac{\epsilon+x_i}{\eta_{ij}^{1/2}}\right)\right]\\\notag
=&1-E\left\{\left[1-E\left\{\frac{1}{\pi}\tan^{-1}\left(\frac{\epsilon-x_i}{\eta_{ij}^{1/2}}\right) + \frac{1}{\pi}\tan^{-1}\left(\frac{\epsilon+x_i}{\eta_{ij}^{1/2}}\right)\right\}\right]^{|T_i|}\right\}\\\notag
=&1-\left[1-\gamma + \gamma\left\{1-E\left\{\frac{1}{\pi}\tan^{-1}\left(\frac{\epsilon-x_i}{\eta_{ij}^{1/2}}\right)+\frac{1}{\pi}\tan^{-1}\left(\frac{\epsilon+x_i}{\eta_{ij}^{1/2}}\right)\right\}\right\}\right]^M\\\notag
=&1-\left[1-\gamma E\left\{\frac{1}{\pi}\tan^{-1}\left(\frac{\epsilon+x_i}{\eta_{ij}^{1/2}}\right)-\frac{1}{\pi}\tan^{-1}\left(\frac{x_i-\epsilon}{\eta_{ij}^{1/2}}\right)\right\}\right]^M
\end{align}

Note that $\tan^{-1}(z+\epsilon) - \tan^{-1}(z-\epsilon)\leq  2\tan^{-1}{\epsilon}\leq 2\epsilon$, for $\epsilon\geq 0$. Therefore,
\begin{align}\notag
&\mathbf{Pr}\left(\left|\hat{x}_{i,min,\gamma}\right|\leq   \epsilon, x_i \neq 0\right)\\\notag
=&1-\left[1-\gamma E\left\{\frac{1}{\pi}\tan^{-1}\left(\frac{\epsilon+x_i}{\eta_{ij}^{1/2}}\right)-\frac{1}{\pi}\tan^{-1}\left(\frac{x_i-\epsilon}{\eta_{ij}^{1/2}}\right)\right\}\right]^M\\\notag
\leq&1-\left[1-\frac{2}{\pi}\gamma\tan^{-1}\epsilon\right]^M
\end{align}
which approaches zero as $\epsilon\rightarrow0$. $\hfill\Box$

%
%\section{Minimum Estimator Followed by a Least Square}
%
%Basically, we can re-use the measurements to do a least square on the top-ranked coordinates by $|x_{i,min,\gamma}|$. Because $\log N/\delta$ will be larger than (e.g.,) $2K$, we use re-use all $M$ measurements. This should provide a good solution to the of  problem of noises (either in the signal or in the measurements).

\subsection{The Worst Case Complexity Bound}

From the worst-case false positive probability bound: $\mathbf{Pr}\left(\left|\hat{x}_{i,min,\gamma}\right|>  \epsilon, x_i = 0\right)\leq\left[1-\gamma\left(1-\gamma\right)^K\right]^M$, by choosing $\gamma = 1/K$ (and $\epsilon\rightarrow0$), we can easily obtain the following Theorem regarding the sample complexity of only using the absolute minimum estimator.
\begin{theorem}\label{thm_worst_complexity}
Using the absolute minimum estimator and $\gamma = 1/K$, for perfect support recovery (with probability $>1-\delta$), it suffices to use
\begin{align}
M \geq& \frac{\log N/\delta}{\log\frac{1}{1-\frac{1}{K}\left(1-\frac{1}{K}\right)^K}}\\
\approx& eK\log N/\delta
\end{align}
measurements.
\end{theorem}
\noindent\textbf{Remark:}\hspace{0.2in} The term $\frac{1}{K}\log\frac{1}{1-\frac{1}{K}\left(1-\frac{1}{K}\right)^K}$ approaches $e = 2.7183...$ very quickly. For example, the difference is only 0.1 when $K=10$.

\section{Analysis of the Absolute Minimum Estimator on Ternary Signals}

Although the complexity result in Theorem~\ref{thm_worst_complexity} can be theoretically exciting, we would like to better understand why empirically we only need substantially fewer measurements. In this section, for convenience, we consider the special case of ``ternary'' signals, i.e., $x_i \in \{-1, 0, 1\}$.   The exact expectation (\ref{eqn_fp_exp}), i.e.,
\begin{align}\notag
&\mathbf{Pr}\left(\left|\hat{x}_{i,min,\gamma}\right|> \epsilon, \ x_i=0\right)
=\left[1- \gamma E\left\{\frac{2}{\pi}\tan^{-1}\left(\frac{\epsilon}{\eta_{ij}^{1/2}}\right)\right\}\right]^M
\end{align}
which, in the case of ternary data, becomes
\begin{align}
\eta_{ij} = \sum_{i=1}^N |x_t|^2 r_{tj}\sim Binomial(K,\gamma)
\end{align}
For convenience, we write
\begin{align}
\mathbf{Pr}\left(\left|\hat{x}_{i,min,\gamma}\right|> \epsilon, \ x_i=0\right)
=\left[1- \frac{1}{K}\left(\gamma K\right) E\left\{\frac{2}{\pi}\tan^{-1}\left(\frac{\epsilon}{\eta_{ij}^{1/2}}\right)\right\}\right]^M
 = \left[1-\frac{1}{K}H(\epsilon,K,\gamma)\right]^M
\end{align}
where
\begin{align}
H(\epsilon,K,\gamma) = \left(\gamma K\right) E\left\{\frac{2}{\pi}\tan^{-1}\left(\frac{\epsilon}{\sqrt{Z}}\right)\right\}, \hspace{0.5in} Z \sim Binomial(K,\gamma)
\end{align}
which can  be easily computed numerically for given $\gamma$, $K$, and $M$. In order for $\mathbf{Pr}\left(\left|\hat{x}_{i,min,\gamma}\right|> \epsilon, \ x_i=0\right)\leq \delta$ for all $i$, we should have
\begin{align}
M  \geq \frac{K}{H(\epsilon,K,\gamma)}\log N/\delta
\end{align}

It would be much more convenient if we do not have to worry about all combinations of $\gamma$ and $K$. In fact, we can resort to the well-studied {\em poisson approximation} by considering $\lambda = \gamma K$ and defining
\begin{align}
h(\epsilon,\lambda) =& \lambda E\left\{\frac{2}{\pi}\tan^{-1}\left(\frac{\epsilon}{\sqrt{Z}}\right)\right\}, \hspace{0.5in} Z \sim Poisson(\lambda)\\\notag
=&\lambda \sum_{k=0}^\infty \left\{\frac{2}{\pi}\tan^{-1}\left(\frac{\epsilon}{\sqrt{k}}\right)\right\} \frac{e^{-\lambda}\lambda^k}{k!}\\
=&\lambda e^{-\lambda}+\lambda e^{-\lambda} \sum_{k=1}^\infty \left\{\frac{2}{\pi}\tan^{-1}\left(\frac{\epsilon}{\sqrt{k}}\right)\right\} \frac{\lambda^k}{k!}
\end{align}

\begin{figure}[h!]
\begin{center}
\mbox{
\includegraphics[width=3.3in]{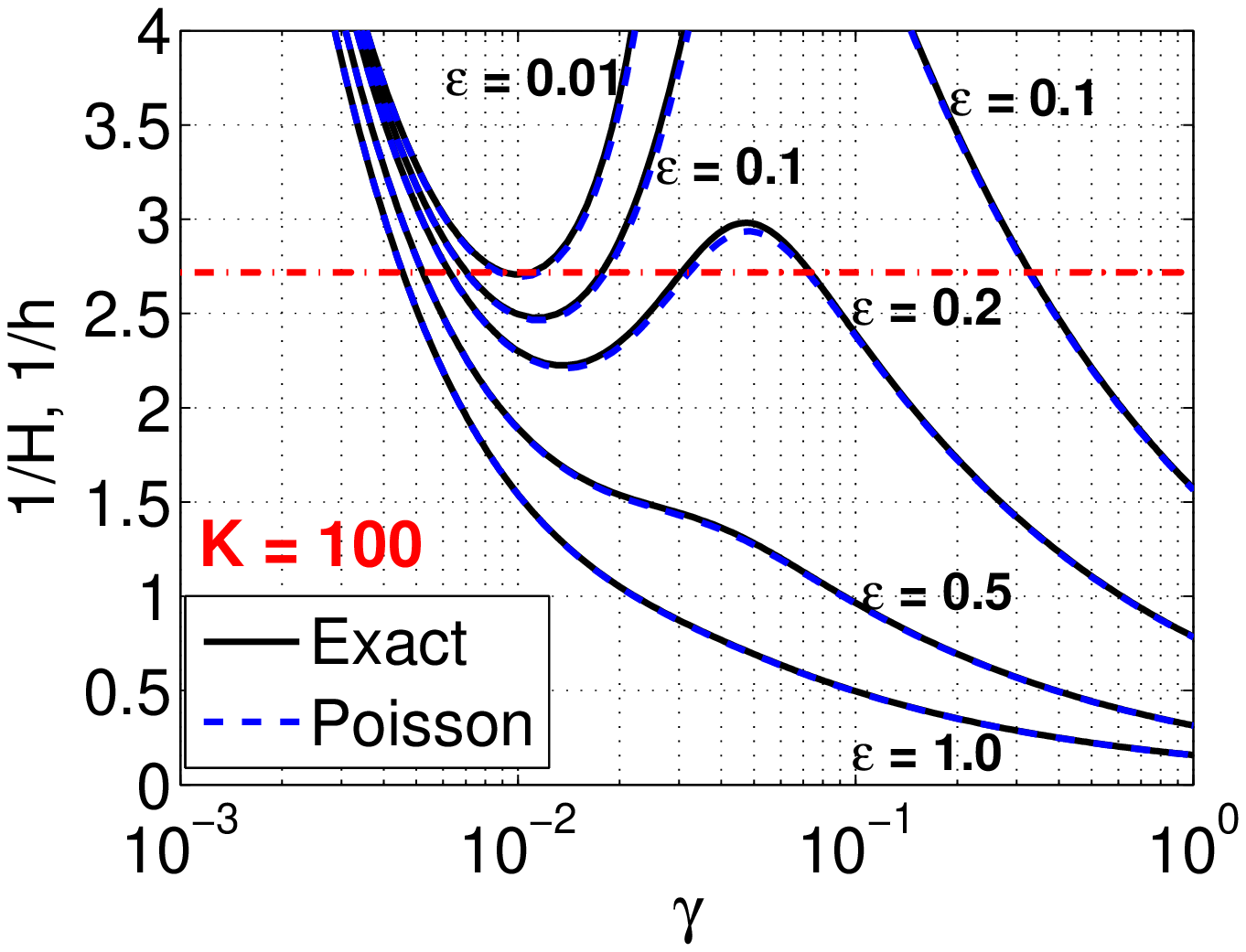}
\includegraphics[width=3.3in]{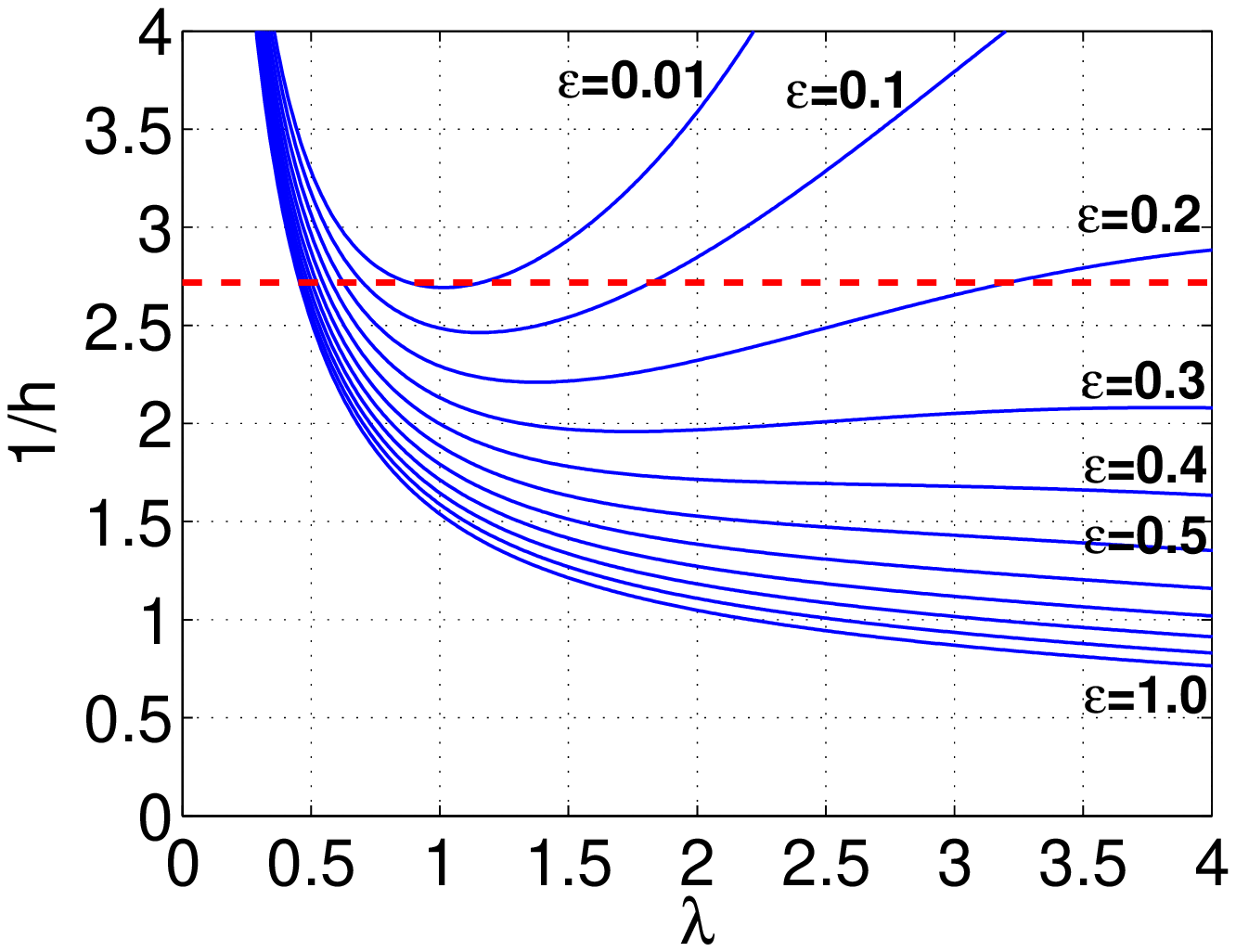}
}
\end{center}
\vspace{-0.2in}
\caption{\textbf{Left Panel}: $\frac{1}{H(\epsilon,K,\gamma)}$ (solid) and $\frac{1}{h(\epsilon,\lambda)}$ (dashed), for $K=100$ and $\epsilon \in \{0.01, 0.1, 0.2, 0.5, 1.0\}$. This plot confirms that the Poisson approximation is indeed very accurate (as expected).
\textbf{Right Panel}: Poisson approximation $\frac{1}{h(\epsilon,\lambda)}$ for $\epsilon\in \{0.01, 0.1, 0.2, 0.3, 0.4, 0.5, 0.6, 0.7, 0.8, 0.9, 1.0\}$. In both panels, we use the horizontal line to indicate $e=2.7183...$. When $\gamma = 1/K$, i.e., $\lambda = 1$, both $\frac{1}{H(\epsilon,K,\gamma)}$  and $\frac{1}{h(\epsilon,\lambda)}$ are upper bounded by $e$.
 }\label{fig_hH}
\end{figure}

Figure~\ref{fig_hH} plots $\frac{1}{H(\epsilon,K,\gamma)}$  and $\frac{1}{h(\epsilon,\lambda)}$ to confirm that the Poisson approximation is very accurate (as one would  expect). At $\gamma = 1/K$ (i.e., $\lambda =1$), the two terms $\frac{1}{H(\epsilon,K,\gamma)}$  and $\frac{1}{h(\epsilon,\lambda)}$ are upper bounded by $e$.  However, when $\epsilon$ is not too small, the constant $e$ can be  conservative. Basically, the choice of $\epsilon$ reflects the level of prior information about the signal. If the signals are significantly away from 0, then we can choose a larger $\epsilon$ and hence the algorithm would require less measurements. For example, if we know the signals are ternary, we can perhaps choose $\epsilon=0.5$ or larger. Also, we can notice that $\gamma = 1/K$ is not necessarily the optimum choice for a given $\epsilon$. In general, the performance is not too sensitive to the choice $\gamma = \lambda/K$ as long  as $\epsilon$ is not too small and the $\lambda$ is reasonably large. This might be good news for practitioners.

\section{Analysis of the Absolute Minimum Estimator with Measurement Noise}

We can also analyze the absolute minimum estimator when measurement noise is present, i.e.,
\begin{align}
\tilde{y}_j = y_j + n_j  = \sum_{i=1}^N x_i \left[s_{ij}r_{ij}\right] + n_j, \hspace{0.2in}\text{where } n_j \sim N(0,\sigma^2), \hspace{0.25in} j = 1, 2, ..., M
\end{align}
Again, we compute the ratio statistic
\begin{align}
\left.\frac{y_j+n_j}{s_{ij}r_{ij}}\right|_{r_{ij}=1}  = \frac{\sum_{t=1}^Nx_ts_{tj}r_{tj} + n_j}{s_{ij}}=x_i + \frac{\sum_{t\neq i}^Nx_ts_{tj}r_{tj}+n_j}{s_{ij}} = x_i + \left(\tilde{\eta}_{ij}\right)^{1/2} \frac{S_2}{S_1}
\end{align}
where $S_1, S_2\sim N(0,1)$, i.i.d., and
\begin{align}
\tilde{\eta}_{ij} = \sum_{t\neq i}^N \left|x_t r_{tj}\right|^2+\sigma^2= \sum_{t\neq i}^N \left|x_t\right|^2 r_{tj}+\sigma^2
\end{align}
\begin{lemma}\label{lem_Pr_fpn}
\textbf{Data-dependent bound:}
\begin{align}
\mathbf{Pr}\left(\left|\hat{x}_{i,min,\gamma}\right|>  \epsilon, x_i = 0\right)=&\left[1- \gamma E\left\{\frac{2}{\pi}\tan^{-1}\left(\frac{\epsilon}{\tilde{\eta}_{ij}^{1/2}}\right)\right\}\right]^M\\
\leq&\left[1- \gamma \left\{\frac{2}{\pi}\tan^{-1}\left\{\frac{{\epsilon}}{\left(\sigma^2+\gamma\sum_{t} x_{t}^2\right)^{1/2}}\right\}\right\}\right]^M
\end{align}
\textbf{Data-independent bound:}
\begin{align}
&\mathbf{Pr}\left(\left|\hat{x}_{i,min,\gamma}\right|> \epsilon, \ x_i=0\right)
\leq\left[1- \gamma \left\{\frac{2}{\pi}\tan^{-1}\left(\frac{\epsilon}{\sigma}\right)\right\}(1-\gamma)^K\right]^M
\end{align}
\textbf{Data-independent complexity bound:}\ \  \ With $\gamma = 1/K$, in order to achieve  $\mathbf{Pr}\left(\left|\hat{x}_{i,min,\gamma}\right|> \epsilon, \ x_i=0\right)\leq \delta$ for all $i$, it suffices to use
\begin{align}
M \geq e \left\{\frac{2}{\pi}\tan^{-1}\left(\frac{\epsilon}{\sigma}\right)\right\} K \log N/\delta
\end{align}
measurements.
\end{lemma}

\noindent\textbf{Proof of Lemma~\ref{lem_Pr_fpn}:} \ \ \
\begin{align}\notag
\mathbf{Pr}\left(\left|\hat{x}_{i,min,\gamma}\right|>  \epsilon, x_i = 0\right)=&\left[1- \gamma E\left\{\frac{2}{\pi}\tan^{-1}\left(\frac{\epsilon}{\tilde{\eta}_{ij}^{1/2}}\right)\right\}\right]^M\\\notag
\leq&\left[1- \gamma \left\{\frac{2}{\pi}\tan^{-1}\left(\frac{{\epsilon}}{\left(E\tilde{\eta}_{ij}\right)^{1/2}}\right)\right\}\right]^M \hspace{0.3in} (\text{Jensen's Inequality})\\\notag
=&\left[1- \gamma \left\{\frac{2}{\pi}\tan^{-1}\left\{\frac{{\epsilon}}{\left(\sigma^2+\gamma\sum_{t} x_{t}^2\right)^{1/2}}\right\}\right\}\right]^M
\end{align}
which is still expressed in terms of the summary of the signal. To obtain a data-independent bound, we have
\begin{align}\notag
&\mathbf{Pr}\left(\left|\hat{x}_{i,min,\gamma}\right|> \epsilon, \ x_i=0\right)
=\left[1- \gamma E\left\{\frac{2}{\pi}\tan^{-1}\left(\frac{\epsilon}{\tilde{\eta}_{ij}^{1/2}}\right)\right\}\right]^M
\leq\left[1- \gamma \left\{\frac{2}{\pi}\tan^{-1}\left(\frac{\epsilon}{\sigma}\right)\right\}(1-\gamma)^K\right]^M
\end{align}
$\hfill\Box$

\vspace{-0.1in}
\section{Analysis of the Tie Estimator}

To construct the {\em tie estimator}, we first compute $z_{ij} = \frac{y_j}{s_{ij}r_{ij}}$ which is anyway needed for the absolute minimum estimator. At each $i$ of interest, we sort those $M$ $z_{ij}$ values and examine the order statistics, $z_{i,(1)}\leq z_{i,(2)}\leq ... \leq z_{i,(M)}$, and their consecutive differences, $z_{i,(j+1)} - z_{i,(j)}$ for $j = 1, 2, ..., M-1$. Then
\begin{align}\notag
\hat{x}_{i,tie,\gamma} = z_{i,(j_i)}, \hspace{0.2in} \text{if } z_{i,(j_i+1)} - z_{i,(j_i)} = 0, \text{ and } |z_{i,(j_i)}|\neq \infty\\\notag
\end{align}

The analysis of the tie estimator is actually not difficult. Recall
\begin{align}\notag
\left.\frac{y_j}{s_{ij}r_{ij}}\right|_{r_{ij}=1}  = \frac{\sum_{t=1}^Nx_ts_{tj}r_{tj}}{s_{ij}}=x_i + \frac{\sum_{t\neq i}^Nx_ts_{tj}r_{tj}}{s_{ij}} = x_i + \left(\eta_{ij}\right)^{1/2} \frac{S_2}{S_1}
\end{align}
where $S_1, S_2\sim N(0,1)$, i.i.d., and  $\eta_{ij} = \sum_{t\neq i}^N \left|x_t\right|^2 r_{tj}$, which has a certain probability of being zero. If $\eta_{ij} = 0$, then $\left.\frac{y_j}{s_{ij}r_{ij}}\right|_{r_{ij}=1} = x_i$.  To reliably estimate the magnitude of $x_i$, we need $\eta_{ij}=0$ to happen more than once, i.e., there should be a tie. Note that
\begin{align}
\mathbf{Pr}\left(\eta_{ij} = 0, r_{ij} =1\right) = \left\{
\begin{array}{lc}
\gamma(1-\gamma)^K &\text{if } \ x_i = 0\\
\gamma(1-\gamma)^{K-1} &\text{if } \ x_i \neq 0
\end{array}
\right.
\end{align}

For a given nonzero coordinate $i$, we would like to have $\eta_{ij} = 0$ more than once among $M$ measurements. This is a binomial problem, and the error probability is simply
\begin{align}
\left[1-\gamma \left(1-\gamma\right)^{K-1}\right]^M + M\left(\gamma \left(1-\gamma\right)^{K-1}\right)\left[1-\gamma \left(1-\gamma\right)^{K-1}\right]^{M-1}
\end{align}

Suppose we use $\gamma=1/K$. To ensure this error is smaller than $\delta$ for all $K$ nonzero coordinates, it suffices to choose $M$ so that
\begin{align}
K\left\{\left[1-\gamma \left(1-\gamma\right)^{K-1}\right]^M + M\left(\gamma \left(1-\gamma\right)^{K-1}\right)\left[1-\gamma \left(1-\gamma\right)^{K-1}\right]^{M-1}\right\} \leq \delta
\end{align}

It is easy to see that this choice of $M$ suffices for recovering the entire signal, not just the nonzero entries. This is due to the nice property of the tie estimator, which has no false positives. That is, if there is a tie, we know for sure that it reveals the true value of the coordinate. For any zero coordinate, either there is no tie or is the tie zero. Therefore, it suffices to choose $M$ to ensure all the nonzero coordinates are recovered.

\begin{theorem}\label{thm_tie_complexity}
Using the tie estimator and $\gamma = \frac{1}{K}$, for perfect signal recovery (with probability $>1-\delta$), it suffices to choose  the number of measurements to be
\begin{align}
M \geq& 1.551 eK\log K/\delta, \hspace{0.2in} \delta \leq 0.05
\end{align}
\end{theorem}

\vspace{0.1in}

\noindent\textbf{Proof of Theorem~\ref{thm_tie_complexity}}: \hspace{0.1in} The recovery task is trivial when $K=1$. Consider $K\geq 2$ and  $p = \frac{1}{K}\left(1-\frac{1}{K}\right)^{K-1}$, i.e., $p\leq 1/4$.  We need to choose $M$ such that $K\left((1-p)^M + Mp(1-p)^{M-1}\right) \leq \delta$.
 Let $M_1$ be such that $K (1-p)^{M_1} = \delta$, i.e., $M_1 = \frac{\log \delta/K}{\log(1-p)} = \frac{\log K/\delta}{\log\frac{1}{1-p}}$.  Suppose we choose  $M=(1+\alpha) M_1$.  Then.
\begin{align}\notag
K\left((1-p)^{(1+\alpha) M_1} + (1+\alpha) M_1p(1-p)^{(1+\alpha)M_1-1}\right)  = \delta\left( \left(\delta/K\right)^\alpha + (1+\alpha)\frac{\log K/\delta}{\log\frac{1}{1-p}}\frac{\left(\delta/K\right)^\alpha}{1-p} p\right)
\end{align}
Therefore, we need to find the $\alpha$  so that
\begin{align}\notag
T(\delta, K, \alpha) = \left(\delta/K\right)^\alpha + \frac{(1+\alpha)\log (K/\delta) \left(\delta/K\right)^\alpha }{\log(1-p)(1-1/p)}\leq 1
\end{align}
Since $p\leq 1/4$,  we have $\frac{\partial}{\partial p}\log(1-p)(1-1/p) = \left(\log(1-p)+p\right)/p^2<0$. Because $p$ is decreasing in $K$, we know that $\frac{1}{\log(1-p)(1-1/p)}$ is decreasing in $K$.  Also, note that
\begin{align}\notag
&\frac{\partial}{\partial K} \left[\log (K/\delta) \left(\delta/K\right)^\alpha\right] = \left(\delta/K\right)^\alpha/K \left(1- \alpha\log K/\delta\right)\\\notag
&\frac{\partial}{\partial \delta } \left[\log (K/\delta) \left(\delta/K\right)^\alpha\right] = \left(\delta/K\right)^\alpha/\delta \left(-1+ \alpha\log K/\delta\right)
\end{align}
As we consider $K\geq 2$ and $\delta \leq 0.05$, we know that, as long as $\alpha \geq 1/\log \frac{2}{0.05} = 1/\log 40$, the term $\log K/\delta \left(\delta/K\right)^\alpha$ is  increasing in $\delta$ and  decreasing in $K$.   Combining the calculations, we know that $T(\delta, K,\alpha)$ is decreasing in $K$ and increasing in $\delta$, for $\alpha>1/\log 40$. It is thus suffices to consider $\delta  = 0.05$ and $K=2$. Because $T(0.05,2,\alpha)$ is  decreasing in $\alpha$, we only need to numerically find the $\alpha$ so that $T(0.05,2,\alpha)=1$, which happens to be $0.5508..$.

Therefore, it  suffices to choose $M = 1.551 M_1 = 1.551\frac{\log K/\delta}{\log\frac{1}{1-\frac{1}{K}\left(1-\frac{1}{K}\right)^{K-1}}}$ measurements. It remains to  show that $\frac{1}{K\log\frac{1}{1-\frac{1}{K}\left(1-\frac{1}{K}\right)^{K-1}}} \leq e$. Due to $\log\frac{1}{1-x}\geq x$, $\forall\ 0<x<1$, we have
\begin{align}\notag
\frac{1}{K\log\frac{1}{1-\frac{1}{K}\left(1-\frac{1}{K}\right)^{K-1}}}\leq \frac{1}{K} \frac{1}{\frac{1}{K}\left(1-\frac{1}{K}\right)^{K-1}} = \frac{1}{\left(1-\frac{1}{K}\right)^{K-1}} = \left(1+\frac{1}{K-1}\right)^{K-1} \leq e
\end{align}
$\hfill\Box$

\section{An Experimental Study}

Compressed sensing is an important problem of broad interest, and it is crucial to experimentally verify that the proposed method performs well as predicted by our theoretical analysis. In this study, we closely  follow the experimental setting as in the well-known wiki page (see~\cite{Url:wiki_sparse}), which compared count-min sketch, SMP, and L1 decoding, on ternary (i.e., $\{-1, 0, 1\}$) signals. In particular, the results for $N=20000$ are available for all three algorithms.  Their results have shown that, in order to achieve similar recovery accuracies,  count-min sketch needs about $10$ to 15 times more measurements than L1 decoding and SMP only needs about half of the measurements of count-min sketch.\\

\begin{figure}[h!]
\begin{center}
\mbox{
\includegraphics[width = 3.45in]{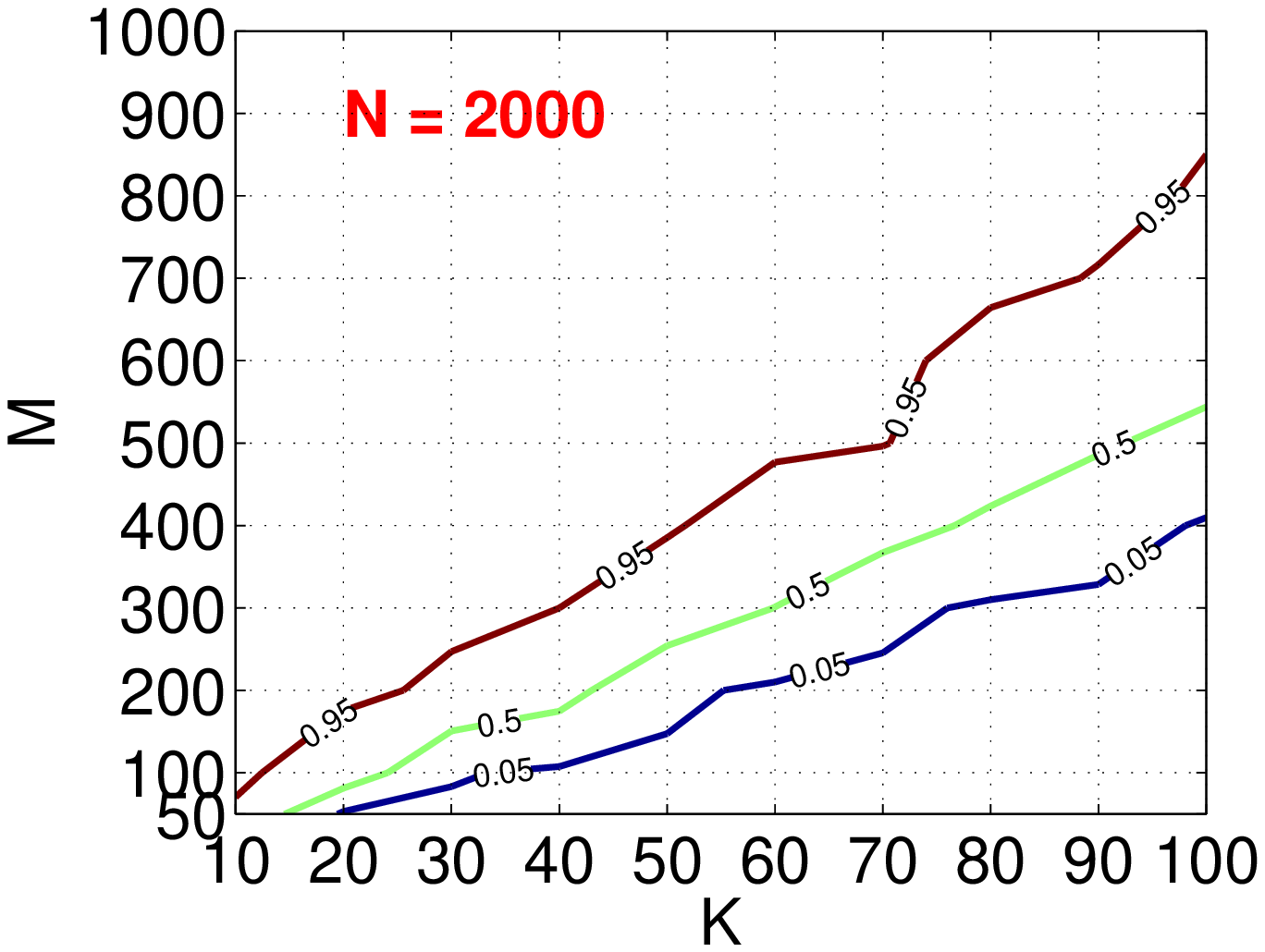}\hspace{-0.15in}
\includegraphics[width = 3.45in]{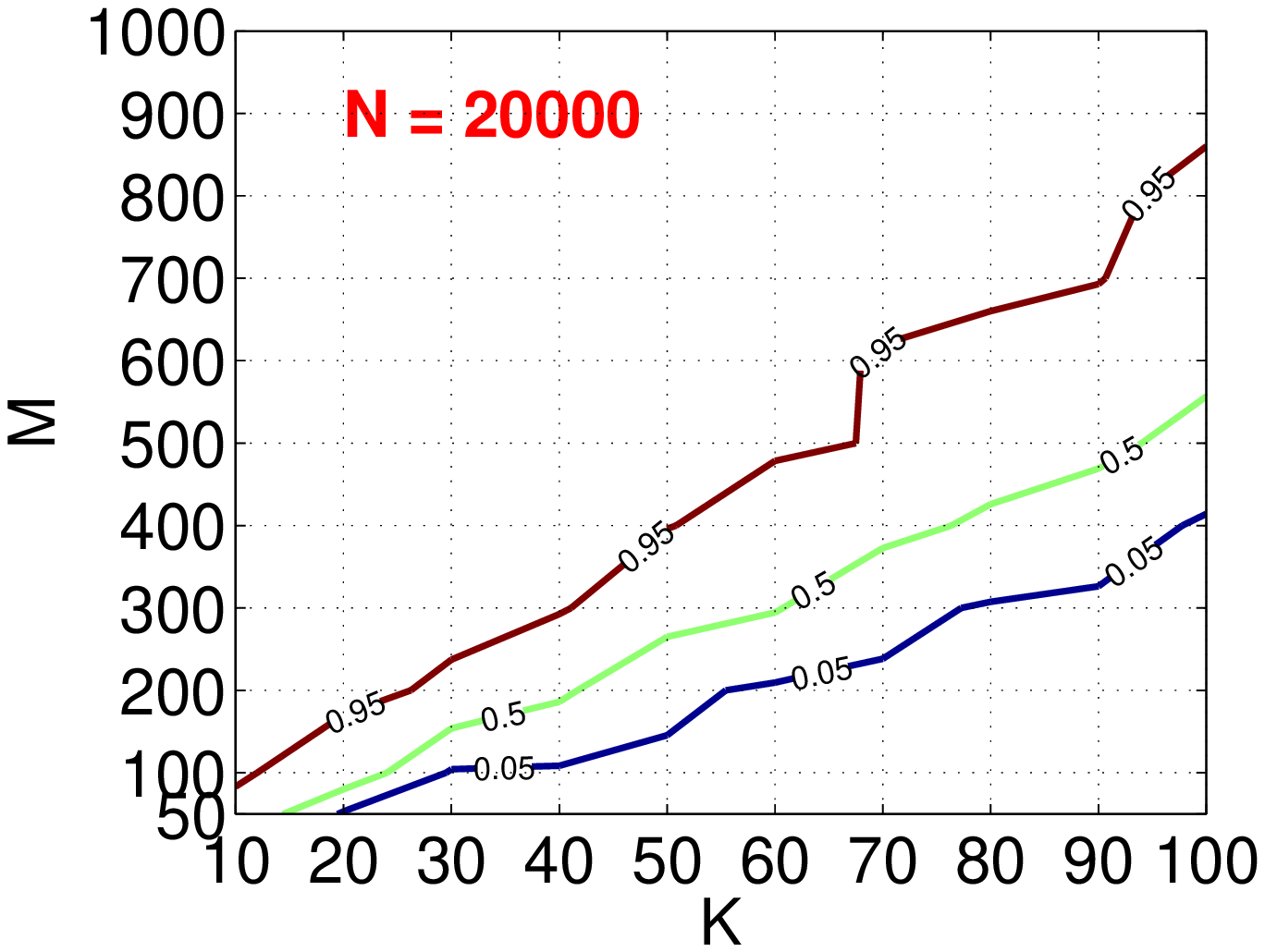}
}
\mbox{
\includegraphics[width = 3.45in]{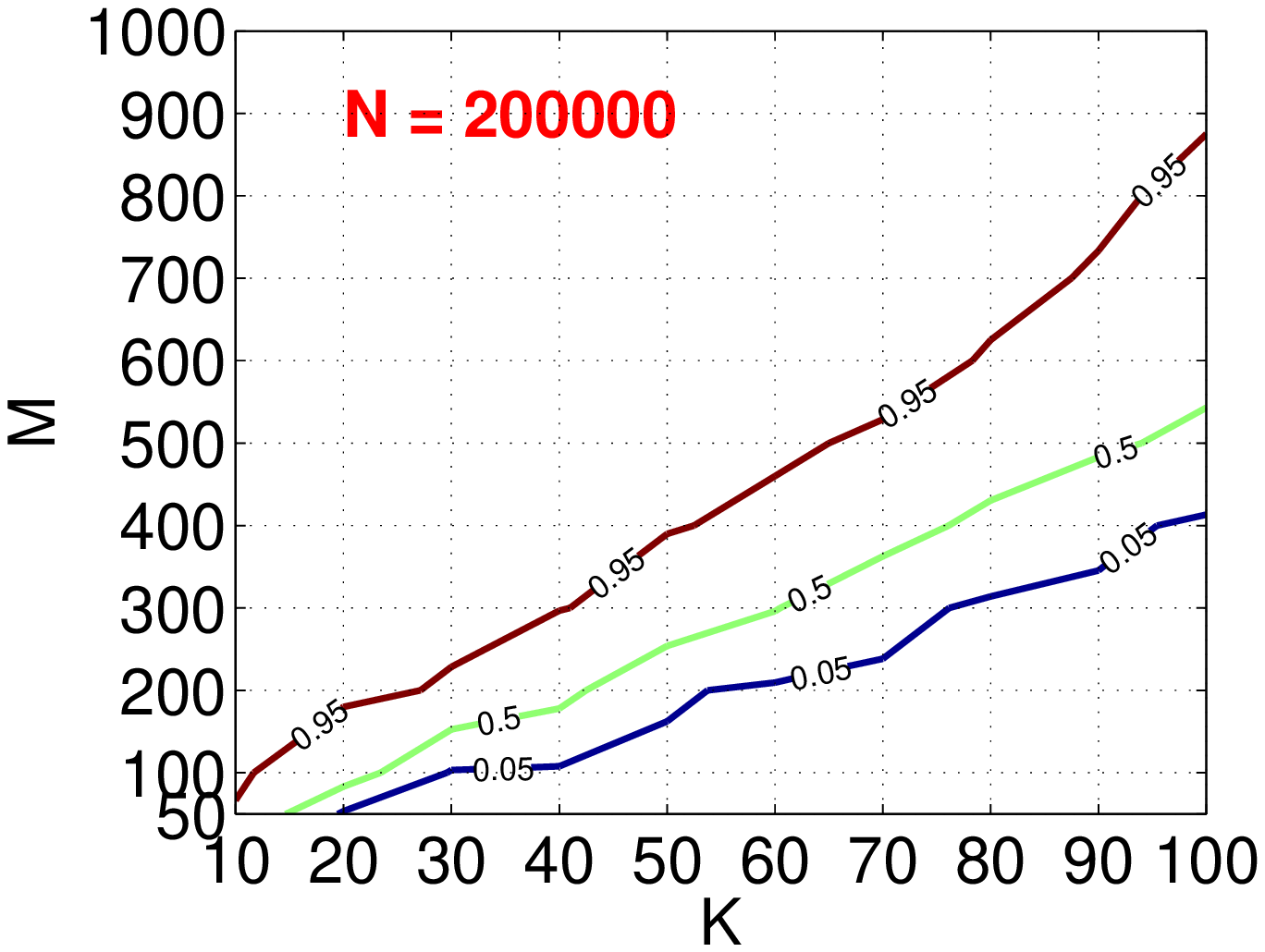}\hspace{-0.15in}
\includegraphics[width = 3.45in]{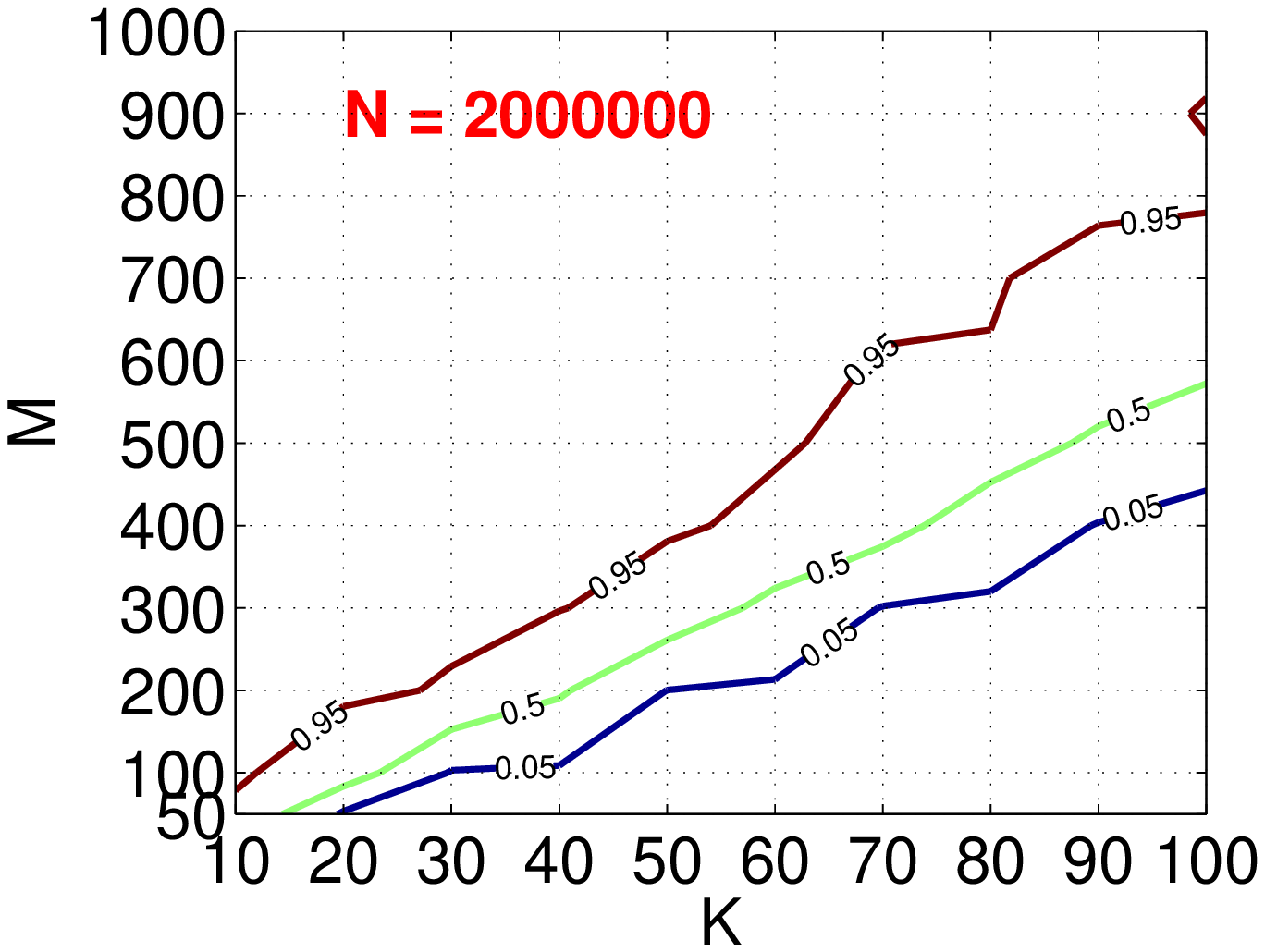}

}
\end{center}
\vspace{-0.2in}
\caption{Contour plot of the empirical success probabilities of our proposed method, for $N=2000$, $20000$, $200000$, and $2000000$. For each combination  $(N, M, K)$, we repeated the simulation 100 times. For $N=20000$, we can see from the wiki page~\cite{Url:wiki_sparse} that our prosed method  provides accurate recovery  results compared to L1 decoding.}\label{fig_g1s1}
\end{figure}

As shown in the success probability contour plot  in Figure~\ref{fig_g1s1} (for $\gamma = 1/K$), the accuracy of our proposed method is (at least) similar to    the accuracy of L1 decoding (based on~\cite{Url:wiki_sparse}).  This should be exciting because, at the same number of measurements, the decoding cost of our proposed algorithm is roughly the same as count-min sketch.

\section{Conclusion}

Compressed sensing has become a popular and important research topic. Using a sparse design matrix has a significant advantage over dense design. For example, in sensing networks, we can replace a dense constellation of sensors by a randomly sparsified one, which may result in substantially saving of sensing hardware and labor costs. In this paper, we show another advantage from the computational perspective of the decoding step. It turns out that using a very sparse design matrix can lead to a computationally very efficient recovery algorithm  without losing accuracies (compared to L1 decoding).

%\bibliographystyle{plain}
%\bibliography{../bib/IEEEabrv,../bib/mybibfile}

\end{document}